%Paper: astro-ph/9307004
%From: mannheim@main.phys.uconn.edu (Philip Mannheim)
%Date: Fri, 2 Jul 93 11:56:01 EDT

\magnification=\magstep1
 \baselineskip=0.821truecm
\centerline {\bf LINEAR POTENTIALS AND GALACTIC ROTATION CURVES - FORMALISM}
\vskip 1.90truecm
\centerline {\bf Philip D. Mannheim}
%\vskip 0.30truecm
\centerline {Department of Physics}
\centerline {University of Connecticut}
\centerline {Storrs, CT 06269-3046}
\smallskip
\centerline{mannheim@uconnvm.bitnet}
\vskip 3.00truecm
\centerline {\bf Abstract}
\vskip 0.60truecm
In the first paper in this series we presented a typical set of
galactic rotation curves associated with the conformal invariant
fourth order theory of gravity which has recently been advanced
by Mannheim and Kazanas as a candidate alternative to the standard
second order Einstein theory. Reasonable agreement with data was
obtained for four representative galaxies without the need for any
non-luminous or dark matter. In this second paper we present the
associated formalism. Additionally, we discuss the status of the
Tully-Fisher relation in our theory, compare and contrast our theory
with the dark matter theory, and make some general observations
regarding the systematics of galactic rotation curve data.
\vskip 3.00truecm
$~~~~~~~~~~~~~~$June, 1993$~~~~~~~~~~~~~~~~~~~~~~~~~~~~~~$UCONN-93-5
\vfill\eject
%\baselineskip=0.75 truecm
\hoffset=0.0truein
%\voffset=0.75truein
\hsize=6.5 truein
\noindent
{\bf (I) Introduction}
\medskip

Despite the overwhelming consensus in the community favoring the existence of
dark matter, as of today no explicit astrophysical or elementary particle dark
matter candidate has yet been found, nor has any dark matter flux yet been
measured. While this has not deterred the bulk of the community in any way,
nonetheless a few authors have begun to question the validity of the
underlying Newton-Einstein gravitational theory itself, taking the view that
the need for dark matter in the standard theory may be symptomatic of
an actual breakdown of the standard picture. Since this apparent
need for dark matter is manifest on essentially every single distance scale
from galactic all the way up to cosmological, while no such need is generally
manifest on the much shorter distance scales where the standard theory was
originally established in the first place, it is thus natural to consider the
possibility that new physics (one might even refer to it as dark physics) may
be opening up on these bigger distance scales. Indeed, rather than
interpreting essentially every single
current large distance scale gravitational observation as yet further evidence
for the existence of dark matter (the common practice in both the learned
and the popular literature), these selfsame data can just as equally be
regarded as signaling the repeated failure of the standard theory;
and definitively so if the only matter which actually exists in the Universe is
that which is luminously observable. Thus the psychologically unwelcome
empirical possibility suggested by the data is that Newton's Law of Gravity
may not be the correct weak gravitational theory, and that, accordingly,
Einstein gravity may not then be the correct covariant one.

Now of course both the Newton and Einstein theories enjoy many successes
(enough to convince most people that they are no longer even challengeable at
all), and thus any alternate theory of gravity must be able to recover all
their established features. To achieve this, one way to proceed is to begin
with galactic rotation curve data (perhaps the most clear cut and well explored
situation
where the Newton-Einstein theory demands dark matter) and try to extract out
a new weak gravity limit which encompasses Newton in an appropriate limit
(see e.g. Milgrom (1983) and Sanders (1990)) with a view to then subsequently
working upwards to a covariant generalization (a program which is still in
progress
- see Sanders (1990) for a recent review). However, in order to ensure
encompassing the
Einstein successes from the outset, there is also much merit in beginning
covariantly and then working downwards to a weak gravity limit. This latter
approach has been advanced and explored by Mannheim and Kazanas in a recent
series of papers (Mannheim and Kazanas (1989, 1991, 1992), Mannheim (1990,
1992, 1993a, 1993b), Kazanas (1991), Kazanas and Mannheim (1991a, 1991b)).
Noting that there is currently no known theoretical reason which would
select out the standard second order Einstein theory from amongst the infinite
class of (all order) covariant, metric based theories of gravity that one
could in principle at least consider, Mannheim and Kazanas reopened the
question of what the correct covariant theory of gravity might be and
developed an approach which works down from an additional fundamental
principle above and beyond covariance, namely that of local scale or conformal
invariance, i.e. invariance under any and all local conformal stretchings
$g_{\mu\nu}(x) \rightarrow \Omega^2(x) g_{\mu\nu}(x)$ of the geometry, this
being
the invariance which is now believed to be possessed by the other three
fundamental interactions, the strong, the electromagnetic and the weak. This
invariance forces gravity to be described uniquely by the fourth order action
$$I_W = -\alpha \int d^4x (-g)^{1/2}
C_{\lambda\mu\nu\kappa}C^{\lambda\mu \nu\kappa}
=-2\alpha \int d^4 x (-g)^{1/2}
(R_{\lambda\mu}R^{\lambda\mu} - (R^{\alpha}_{\phantom {\alpha}\alpha})^{2}/3)
\eqno(1) $$
where $C_{\lambda\mu\nu\kappa}$ is the conformal Weyl tensor and $\alpha$
is a purely dimensionless coefficient. In their original paper Mannheim
and Kazanas (1989) obtained the complete and exact, non-perturbative exterior
vacuum solution associated with a static,
spherically symmetric gravitational source such as a star in this theory, viz.
$$-g_{00}= 1/g_{rr}=1-\beta(2-3 \beta \gamma )/r - 3 \beta \gamma
+ \gamma r - kr^2 \eqno(2)$$
where $\beta, \gamma,$ and $k$ are three appropriate dimensionful
integration constants. As can be seen, for small enough values of the linear
and quadratic terms (i.e. on small enough distance scales) the solution
reduces to the familiar Schwarzschild solution of Einstein gravity, with the
conformal theory then enjoying the same successes as the Einstein theory on
those distance scales. On larger distance scales, however, the theory begins to
differ
from the Einstein theory through the linear potential term, and (with the
quadratic term only possibly being important cosmologically, and with both the
$\beta\gamma$ product terms being found to be numerically negligible in the
fits of Mannheim (1993b)) then yields a non-relativistic gravitational
potential
$$V(r)=-\beta/r+ \gamma r/2 \eqno(3)$$
which may be fitted to data whenever the weak gravity limit is applicable.

The conformal theory thus not only generalizes Newton (Eq.
(3)) it also generalizes Schwarzschild (Eq. (2)), and even does so in way
which is then able to naturally recover both the Newton and Schwarzschild
phenomenologies on the appropriate distance scales. Since the conformal theory
recovers the requisite solutions to Einstein gravity on small enough distance
scales (even while never recovering the Einstein Equations themselves -
observation only demands the recovery of the solutions not of the equations),
that fact alone makes the theory indistinguishable from and just as viable as
the Einstein theory on those distance scales, something recognized by
Eddington (1922) as far back as the very early days of Relativity. (Eddington
was not
aware of the full exact solution of Eq. (2) but was aware that it was a
solution
to fourth order gravity in the restricted case where $\gamma=0$. It was only
much
later that the complete and exact solution of Eq. (2) was found and that its
consistency
was established by successfully matching it on to the associated
exact interior solution (Mannheim and Kazanas (1992)). Thus in this sense
conformal gravity should always have been considered as a viable explanation
of solar system physics. That it never was so considered was in part due to
the fact that strict conformal symmetry requires that all particles be
massless, something which would appear to immediately rule the symmetry out.
However,
with the advent of modern spontaneously broken gauge theories manifest in the
other three fundamental interactions, it is now apparent that mass can still
be generated in the vacuum in otherwise dimensionless theories like the one
associated with the action of Eq. (1). (And, interestingly, such dynamical mass
generation is even found to still lead to geodesic motion (Mannheim (1993a)),
despite the fact that the associated mass generating Higgs field
which accompanies a test particle carries its own energy and momentum which the
gravitational field also sees). Hence, it would appear that today the only
non-relativistic way to distinguish between the two covariant theories is to
explore
their observational implications on larger distance scales
where the linear potential term first makes itself manifest.

A first step towards this phenomenological end was taken recently by Mannheim
(1993b) with the above non-relativistic potential $V(r)$ being used in
conjunction with
observed surface brightness data to fit the rotation
curves of four representative galaxies. The particular choice of galaxies was
guided by
the recent comprehensive survey of the $HI$ rotation curves of spiral galaxies
made by
Casertano and van Gorkom (1991) who found that those data fall into essentially
four general groups characterized by specific correlations between the maximum
rotation
velocity and the luminosity. In order of increasing luminosity the four groups
are
dwarf, intermediate, compact bright, and large bright galaxies. Thus one
representative
galaxy from each group was studied, respectively the galaxies DDO154 (a gas
dominated
rather than star dominated galaxy), NGC3198, NGC2903, and NGC5907, with the
fitting of
Mannheim (1993b) being reproduced here as Fig. (1). (The reader is referred to
the
original paper for details). For NGC3198 the rotation curve of Begeman (1989)
and the
surface brightness data of Wevers et al. (1986) and Kent (1987) were used, for
NGC 2903
the data were taken from Begeman (1987) and Wevers et al. (1986), for NGC 5907
from
van Albada and Sancisi (1986) and Barnaby and Thronson (1992), and for DDO154
from
Carignan and Freeman (1988) and Carignan and Beaulieu (1989). (While Carignan
and his
coworkers favor a distance of 4 Mpc to DDO154, Krumm and Burstein (1984) favor
10 Mpc. Since the gas contribution is extremely distance sensitive, for
completeness we
opted to fit this galaxy at both the candidate distances). As can be seen from
Fig. (1),
the conformal theory appears to be able to do justice to a data set which
involves a broad
range of luminosities, and to even do so without the need for dark matter, a
point we
analyze further below.

In order to apply the linear potential to an extended object such as a disk
it was found helpful to develop a general formalism, with the results of the
formalism
being used in Mannheim (1993b) to produce the fits of Fig. (1). In the present
paper we
present the actual details of the derivation of the formalism (something that
will be
useful for future studies), with the formalism actually even being
of interest in its own right since it extends to linear potentials the earlier
work of
Toomre (1963), Freeman (1970), and Casertano (1983) on Newtonian disks. We
present
the general formalism in Sec. (2), and in Sec. (3) we present some general
comments
on our work, discuss the status of the Tully-Fisher (Tully and Fisher (1977))
relation in conformal gravity, and compare and contrast our theory
with the standard dark matter theory.

\medskip
\noindent
{\bf (2) The Potential of an Extended Disk}
\medskip

In order to handle the weak gravity potential of an extended object such as a
disk
of stars each with gravitational potential $V(r)=-\beta/r+ \gamma r/2 $ many
ways are possible with perhaps the most popular being due to Toomre (1963).
Since the method he developed for the Newtonian case does not immediately
appear to generalize to linear potentials, we have instead generalized his
approach to
non-thin disks (a step also taken by  Casertano (1983)) and
then to disks with linear potentials.
To determine the Newtonian potential of an axially symmetric
(but not yet necessarily thin) distribution of matter sources with
luminosity density function
$\rho(R,z^{\prime})$ we need to evaluate the quantity
$$V_{\beta}(r,z)=-\beta
\int_0^{\infty}dR \int_0^{2\pi}d\phi^{\prime}
\int_{-\infty}^{\infty}dz^{\prime}
{R\rho(R,z^{\prime}) \over
(r^2+R^2-2rRcos\phi^{\prime}+(z-z^{\prime})^2)^{1/2}} \eqno(4)$$
where $R,~\phi^{\prime},~z^{\prime}$ are cylindrical source coordinates
and $r$ and $z$ are the only observation point coordinates of relevance.
To evaluate Eq. (4) it is convenient to make use of the cylindrical
Green's function Bessel function expansion
$${1 \over \vert {\bf r} -{\bf r^{\prime}} \vert }=\sum_{m=-\infty}^{\infty}
\int_0^\infty dk J_m(kr)J_m(kr^{\prime})
e^{im(\phi-\phi^{\prime})-k \vert  z -  z^{\prime} \vert } \eqno(5)$$
whose validity can readily be checked by noting that use of the identity
$$\nabla^2[J_m(kr)e^{im\phi-k \vert  z -  z^{\prime} \vert }]
=-2kJ_m(kr)e^{im\phi}\delta(z-z^{\prime}) \eqno(6)$$
leads to the relation
$$\nabla^2 \left( {1 \over \vert {\bf r} -{\bf r^{\prime}} \vert }\right)=
-4\pi\delta^3({\bf r} -{\bf r^{\prime}}) \eqno (7)$$
(In his original study Toomre used a Bessel function
discontinuity formula (essentially Eq. (6))
which only appears to be applicable to thin disks. Using
the full completeness properties of the Bessel functions enables us to treat
non-thin disks as well).
While Eq. (5) is standard, it is not utilized as often as the more
familiar modified Bessel function expansion
$${1 \over \vert {\bf r} -{\bf r^{\prime}} \vert }={2 \over \pi}
\sum_{m=-\infty}^{\infty}
\int_0^\infty dk cos[k (  z -  z^{\prime})]
I_m(kr_{<})K_m(kr_{>}) e^{im(\phi-\phi^{\prime})}\eqno (8)$$
since the product of the two modified Bessel functions has much better
convergence properties at
infinity than the product of the two ordinary Bessel functions. Nonetheless,
the ordinary Bessel
functions do actually vanish at infinity which is sufficient for our purposes
here. A
disadvantage of the expansion of Eq. (8) is that it
involves oscillating $z$ modes rather than the bounded $z$ modes
given in Eq. (5), with the bounded form of Eq. (5) actually being extremely
convenient
for a disk whose matter distribution is concentrated around $z=0$. An
additional
shortcoming of the expansion of Eq. (8) is that when it is inserted into
Eq. (4) it requires the $R$ integration range to be broken up into two
separate pieces at the point of observation.
However, inserting Eq. (5) into Eq. (4) leads to
$$V_{\beta}(r,z)=-2\pi\beta\int_{0}^{\infty} dk\int_{0}^{\infty}dR
\int_{-\infty}^{\infty}dz^{\prime}
R \rho(R,z^{\prime})J_0(kr)J_0(kR) e^{-k\vert z -  z^{\prime}\vert}\eqno(9)$$
which we see requires no such break up. Finally, taking the disk to be
infinitesimally thin (viz. $\rho(R,z^{\prime})=\Sigma(R)\delta(z^{\prime})$)
then yields for points with $z=0$ the potential
$$V_{\beta}(r)=-2\pi\beta\int_{0}^{\infty} dk\int_{0}^{\infty}dR
R \Sigma(R)J_0(kr)J_0(kR) \eqno(10)$$
which we immediately recognize as Toomre's original result for an
infinitesimally thin disk. In passing we note that Eq. (9) also holds for
points
which do not lie in the $z=0$ plane of the disk, and also applies to disks
whose
thickness may not in fact be negligible, with the form of Eq. (9) being
particularly convenient if the fall-off of the matter distribution in the
$z$ direction is itself exponential (see below).

For our purposes here, the expansion of Eq. (5) can immediately be
applied to the linear potential case too, and this leads directly
(on setting $\vert {\bf r} -{\bf r^{\prime}} \vert =
({\bf r} -{\bf r^{\prime}})^2/
\vert {\bf r} -{\bf r^{\prime}} \vert$) to the potential
$$V_{\gamma}(r,z)={\gamma \over 2}
\int_0^{\infty}dR \int_0^{2\pi}d\phi^{\prime}
\int_{-\infty}^{\infty}dz^{\prime}
R\rho(R,z^{\prime})
[r^2+R^2-2rRcos\phi^{\prime}+(z-z^{\prime})^2]^{1/2}$$
$$=\pi\gamma
\int_0^{\infty}dk \int_0^{\infty}dR \int_{-\infty}^{\infty}dz^{\prime}
R\rho(R,z^{\prime})
[(r^2+R^2+(z-z^{\prime})^2)J_0(kr)J_0(kR)$$
$$-2rR J_1(kr)J_1(kR)]
e^{-k\vert z -  z^{\prime}\vert} \eqno(11)$$
Equation (11) then  reduces at $z=0$ for infinitesimally thin disks to the
compact expression
$$V_{\gamma}(r)=\pi\gamma\int_{0}^{\infty} dk\int_{0}^{\infty}dRR
\Sigma(R)[ (r^2+R^2)J_0(kr)J_0(kR) -2rR J_1(kr)J_1(kR)]
\eqno(12)$$

If the $k$ integrations are performed first in Eqs. (10) and (12) they
lead to singular hypergeometric functions whose subsequent $R$
integrations contain infinities which, even while they are in fact mild enough
to be
integrable (as long as $\Sigma(R)$ is sufficiently damped at infinity),
nonetheless require a little care when being carried out numerically.
Thus unlike the sphere whose potential is manifestly finite at every step of
the
calculation, the disk, because of its lower dimensionality, actually encounters
infinities
at any interior point of observation on the way to a final finite answer.
However, since
the final
answer is finite, it should be possible to obtain this answer without ever
encountering
any infinities at any stage of the calculation at all; and indeed, if the
distribution
function $\Sigma(R)$ is available in a closed form, then performing the $R$
integration before the $k$ integration can yield a calculation which is finite
at every
stage. Thus, for the exponential disk
$$\Sigma(R)=\Sigma_0e^{-\alpha R} \eqno (13)$$
where $1/\alpha=R_0$ is the scale length of the disk and $N=2\pi\Sigma_0 R_0^2$
is the number of stars in the disk, use of the standard Bessel
function integral formulas
$$\int_0^\infty dR RJ_0(kR)e^{-\alpha R}={\alpha \over
(\alpha^2+k^2)^{3/2}} \eqno(14)$$
$$\int_0^\infty dk {J_0(kr) \over (\alpha^2+k^2)^{3/2}}
  =(r/2\alpha)[I_0(\alpha r/2)K_1(\alpha r/2)-
I_1(\alpha r/2)K_0(\alpha r/2)]
\eqno(15)$$
then leads directly to Freeman's original result, viz.

$$V_{\beta}(r)=-2\pi\beta\Sigma_0 \int_{0}^{\infty}dk
{\alpha J_0(kr) \over (\alpha^2+k^2)^{3/2}}$$
$$= -\pi\beta\Sigma_0 r[I_0(\alpha r/2)K_1(\alpha r/2)-
I_1(\alpha r/2)K_0(\alpha r/2)]\eqno(16)$$
for the Newtonian potential of an exponential disk. The use of the additional
integral
formula
$$\int_0^\infty dR R^2J_1(kR)e^{-\alpha R}={3\alpha k\over
(\alpha^2+k^2)^{5/2}} \eqno (17)$$
and a little algebra (involving eliminating $J_1(kr)=-(dJ_0(kr)/dk)/r$ via
an integration by parts) enable us to obtain for the
linear potential contribution the expression
$$V_{\gamma}(r)=
\pi\gamma\Sigma_0
\int_{0}^{\infty}dk
\left( {\alpha r^2 J_0(kr) \over (\alpha^2+k^2)^{3/2}}
-{9\alpha J_0(kr) \over (\alpha^2+k^2)^{5/2}}
+{15\alpha^3 J_0(kr) \over (\alpha^2+k^2)^{7/2}}
-{6\alpha kr J_1(kr) \over (\alpha^2+k^2)^{5/2}} \right)$$
$$=\pi\gamma\Sigma_0
\int_{0}^{\infty}dk J_0(kr)
\left( {\alpha r^2 \over (\alpha^2+k^2)^{3/2}}
+{15\alpha \over (\alpha^2+k^2)^{5/2}}
-{15\alpha^3 \over (\alpha^2+k^2)^{7/2}} \right)
\eqno(18)$$
\noindent
Equation (18) is readily evaluated through use of the modified
Bessel function recurrence relations
$$ I_0^{\prime}(z)=I_1(z)~~~,~~~I_1^{\prime}(z)=I_0(z)-I_1(z)/z $$
$$ K_0^{\prime}(z)=-K_1(z)~~~,~~~K_1^{\prime}(z)=-K_0(z)-K_1(z)/z
\eqno(19)$$
in conjunction with Eq. (15), and yields
$$V_{\gamma}(r)= \pi\gamma\Sigma_0
\{ (r/\alpha^2)[I_0(\alpha r/2)K_1(\alpha r/2)-
I_1(\alpha r/2)K_0(\alpha r/2)]$$
$$+ (r^2/2\alpha)[I_0(\alpha r/2)K_0(\alpha r/2)+
I_1(\alpha r/2)K_1(\alpha r/2)] \}
\eqno(20)$$

To obtain test particle rotational velocities  we need only differentiate
Eqs. (16) and (20) with respect to $r$. This is readily achieved via
repeated use of the recurrence relations of Eqs. (19) which form a closed
set under differentiation so that higher modified Bessel functions such as
$I_2(\alpha r/2)$ and $K_2(\alpha r/2)$ are not encountered; and the procedure
is
found to yield
$$rV^{\prime}(r)=
(N\beta\alpha^3 r^2/2)[I_0(\alpha r/2)K_0(\alpha r/2)-
I_1(\alpha r/2)K_1(\alpha r/2)]$$
$$+(N\gamma r^2\alpha/2)I_1(\alpha r/2)K_1(\alpha r/2)
\eqno(21)$$
Using the asymptotic properties of the modified Bessel functions we find
that at distances much larger than the scale length $R_0$ Eq. (21) yields
$$rV^{\prime}(r) \rightarrow {N\beta \over  r}+
{N\gamma r \over 2} -{3N\gamma R_0^2\over 4 r} \eqno(22)$$
as would be expected. We recognize the asymptotic Newtonian term to be just
$N\beta/r$ where $N$ is the number of stars in the disk. The quantity
$N\beta$ is usually identified as $MG/c^2$ with $M$ being taken to be the
mass of the disk. For normalization purposes it is convenient to use this
coefficient to define the velocity $v_0=(N\beta/R_0)^{1/2}$, the velocity
that a test particle would have if orbiting a Newtonian point galaxy with the
same total mass at a distance of one scale length. In terms of the convenient
dimensionless parameter $\eta=\gamma R_0^2/\beta$ Eq. (21) then yields for the
rotational velocity $v(r)$ of a circular orbit in the plane of a thin
exponential
disk the exact expression
$$v^2(r)/v_0^2=(r^2\alpha^2/2)[I_0(\alpha r/2)K_0(\alpha r/2)+
(\eta-1)I_1(\alpha r/2)K_1(\alpha r/2)] \eqno(23)$$
an expression which is surprisingly compact. For thin disks then all departures
from the
standard Freeman result are thus embodied in the parameter $\eta$ in the simple
manner
indicated.

Beyond making actual applications to galaxies, a further advantage of
having an exact solution in a particular case is that it can be used to test
a direct numerical evaluation of the galactic potential (which involves
integrable infinities) by also running the program for a model exponential
disk. Also, it is possible to perform the calculation analytically in
various other specific cases. For a thin axisymmetric disk with a Gaussian
surface matter distribution $\Sigma(R)=\Sigma_0$exp$(-\alpha^2R^2)$ and
$N=\pi\Sigma_0/\alpha^2$ stars
(this being a possible model for the sometimes steeper central region
of a galaxy in cases where there may be no spherical bulge) we find for the
complete rotational velocity the expression
$$rV^{\prime}(r)=\pi^{1/2}N\beta\alpha^3 r^2
[I_0(\alpha^2r^2/2)-I_1(\alpha^2r^2/2)]e^{-\alpha^2 r^2/2}$$
$$+(\pi^{1/2}N\gamma \alpha r^2/4) [I_0(\alpha^2r^2/2)+
I_1(\alpha^2r^2/2)]e^{-\alpha^2 r^2/2} \eqno(24)$$
Similarly, for a spherically symmetric matter distribution (the central
bulge region of a galaxy) with radial matter density $\sigma(r)$
and $N=4\pi \int  dr^{\prime}r^{\prime 2}\sigma(r^{\prime})$ stars
we obtain the general expression
$$rV^{\prime}(r)={4\pi\beta \over r}\int_0^r dr^{\prime}\sigma(r^{\prime})
r^{\prime 2}$$
$$+{2\pi\gamma \over 3r}\int_0^r dr^{\prime}\sigma(r^{\prime})
(3r^2r^{\prime 2}-r^{\prime 4})
+{4\pi\gamma r^2\over 3}\int_r^{\infty} dr^{\prime}\sigma(r^{\prime})
r^{\prime } \eqno(25)$$
which can readily be integrated once a particular $\sigma(r)$ is specified.

Beyond the exact expressions obtained above there is one other case of
practical interest
namely that of non-thin but separable disks, a case which can also be greatly
simplified
by our formalism. For such separable disks we set
$\rho(R,z^{\prime})=\Sigma(R)f(z^{\prime})$ where the usually
symmetric thickness function $f(z^{\prime})=f(-z^{\prime})$ is normalized
according to
$$\int_{-\infty}^\infty dz^{\prime}f(z^{\prime})
=2\int_0^\infty dz^{\prime}f(z^{\prime})=1 \eqno(26)$$
\noindent
Recalling that
$$e^{-k\vert z -  z^{\prime}\vert}=\theta(z -  z^{\prime})e^{-k(z -
z^{\prime})}+
\theta(z^{\prime}-z)e^{+k(z -  z^{\prime})} \eqno(27)$$
\noindent
we find that Eqs. (9) and (11) then yield for points with $z=0$
$$V_{\beta}(r)=-4\pi\beta\int_{0}^{\infty} dk\int_{0}^{\infty}dR
\int_0^{\infty}dz^{\prime}R \Sigma(R)f(z^{\prime})
J_0(kr)J_0(kR) e^{-kz^{\prime}}\eqno(28)$$
and
$$V_{\gamma}(r)=2\pi\gamma\int_{0}^{\infty} dk\int_{0}^{\infty}dR
\int_0^{\infty}dz^{\prime}R \Sigma(R)f(z^{\prime})$$
$$\times~[(r^2+R^2+z^{\prime 2})J_0(kr)J_0(kR)
-2rR J_1(kr)J_1(kR)]e^{-kz^{\prime}} \eqno(29)$$
in the separable case. Further simplification is possible if the radial
dependence
is again exponential (viz. $\Sigma(R)=\Sigma_0$exp$(-\alpha R))$ and yields,
following
some algebra involving the use of the recurrence relation
$J_1^{\prime}(z)=J_0(z)-J_1(z)/z$, the expressions
$$rV_{\beta}^{\prime}(r)=2N\beta\alpha^3 r \int_{0}^{\infty}dk
\int_0^{\infty}dz^{\prime}{ f(z^{\prime})e^{-kz^{\prime}}kJ_1(kr)
\over (\alpha^2+k^2)^{3/2}} \eqno(30)$$
and
$$rV_{\gamma}^{\prime}(r)=N\gamma\alpha^3 r \int_{0}^{\infty}dk
\int_0^{\infty}dz^{\prime} f(z^{\prime})e^{-kz^{\prime}}$$
$$\times~ \left( -{4rJ_0(kr) \over (\alpha^2+k^2)^{3/2}}
+{6\alpha^2rJ_0(kr) \over (\alpha^2+k^2)^{5/2}}
-{(r^2+z^{\prime 2})kJ_1(kr) \over (\alpha^2+k^2)^{3/2}}
+{9kJ_1(kr) \over (\alpha^2+k^2)^{5/2}}
-{15\alpha^2kJ_1(kr) \over (\alpha^2+k^2)^{7/2}} \right)
\eqno(31)$$

As regards actual specific forms for $f(z^{\prime})$, two particular ones have
been
identified via the surface photometry of edge on galaxies, one by van der Kruit
and
Searle (1981), and the other by Barnaby and Thronson (1992). Respectively they
are
$$f(z^{\prime})=sech^2(z^{\prime}/z_0)/2z_0 \eqno(32)$$
and
$$f(z^{\prime})=sech(z^{\prime}/z_0)/\pi z_0 \eqno(33)$$
each with appropriate scale height $z_0$. We note that both of these thickness
functions
are falling off very rapidly in the $z^{\prime}$ direction just like the
Bessel function expansion itself of Eq. (5). Consequently,
Eqs. (28) - (31) will now have very good convergence properties.
The thickness function of Eq. (32) is found to lead to rotational velocities of
the form
$$rV^{\prime}_{\beta}(r)=
(N\beta\alpha^3 r^2/2)[I_0(\alpha r/2)K_0(\alpha r/2)-
I_1(\alpha r/2)K_1(\alpha r/2)]$$
$$-N\beta\alpha^3 r\int_0^\infty dk {k^2J_1(kr)z_0\beta(1+kz_0/2) \over
(\alpha^2+k^2)^{3/2}} \eqno (34)$$
where
$$\beta(x)=\int_0^1{t^{x-1} \over (1+t)} \eqno(35) $$
and
$$rV_{\gamma}^{\prime}(r)=N\gamma\alpha^3 r \int_{0}^{\infty}dk
(1-kz_0\beta(1+kz_0/2))$$
$$\times~ \left( -{2rJ_0(kr) \over (\alpha^2+k^2)^{3/2}}
+{3\alpha^2rJ_0(kr) \over (\alpha^2+k^2)^{5/2}}
-{r^2kJ_1(kr) \over 2(\alpha^2+k^2)^{3/2}}
+{9kJ_1(kr) \over 2(\alpha^2+k^2)^{5/2}}
-{15\alpha^2kJ_1(kr) \over 2(\alpha^2+k^2)^{7/2}} \right)$$
$$+N\gamma\alpha^3 r \int_{0}^{\infty}dk{kJ_1(kr)
\over 2(\alpha^2+k^2)^{3/2}}
{d^2 \over dk^2}\left( kz_0\beta(1+{kz_0 \over 2})
\right) \eqno(36)$$
Similarly, the thickness function of Eq. (33) leads to
$$rV^{\prime}_{\beta}(r)={2N\beta\alpha^3 r \over \pi}
\int_0^\infty dk {kJ_1(kr)\beta(1/2+kz_0/2) \over (\alpha^2+k^2)^{3/2}}
\eqno (37)$$
and
$$rV_{\gamma}^{\prime}(r)={N\gamma\alpha^3 r \over \pi} \int_{0}^{\infty}dk
\beta(1/2+kz_0/2)$$
$$\times~ \left( -{4rJ_0(kr) \over (\alpha^2+k^2)^{3/2}}
+{6\alpha^2rJ_0(kr) \over (\alpha^2+k^2)^{5/2}}
-{r^2kJ_1(kr) \over (\alpha^2+k^2)^{3/2}}
+{9kJ_1(kr) \over (\alpha^2+k^2)^{5/2}}
-{15\alpha^2kJ_1(kr) \over (\alpha^2+k^2)^{7/2}} \right)$$
$$-{N\gamma\alpha^3 r \over \pi} \int_{0}^{\infty}dk{kJ_1(kr)
\over (\alpha^2+k^2)^{3/2}}
{d^2 \over dk^2}\left(\beta({1+kz_0 \over 2}) \right)
\eqno(38)$$
\noindent
The great utility of these expressions is that all of the functions of
$\beta(x)$
and their derivatives which appear in Eqs. (34) - (38) converge very rapidly to
their asymptotic values as their arguments increase. Consequently the $k$
integrations in Eqs. (34) - (38) converge very rapidly numerically while
encountering no singularities at all.

As a practical matter, the observed scale heights $z_0$ are usually much
smaller than
any observed scale lengths $R_0$. Consequently the thickness corrections of
Eqs.
(34) - (38) usually only modify the thin disk formula of Eq. (21) in the
central galactic
region, and thus have essentially no effect on the linear potential
contribution.
For the Newtonian term
the corrections of Eqs. (34) and (37) to the Freeman formula tend to reduce the
overall
Newtonian contribution (c.f. the second term in Eq. (34) and Casertano (1983))
and serve to
ensure that the inner rotation curves of Fig. (1)
are well described (see Mannheim (1993b)) by the luminous Newtonian
contribution, to thus
clear the way to explore the effect of the linear term on the outer region of
the
rotation curve, a region where its presence is significant and where the thin
disk formula
of Eq. (21) provides a very good approximation to the dynamics.
\medskip
\noindent
{\bf (3) General Comments}
\medskip
In order to understand the general features of the rotation curves of Fig. (1)
it is
instructive to consider the generic implications of the thin disk formula of
Eq. (23), a two
parameter formula with $v_0$ fixing the overall normalization and $\eta$ the
relative
contributions of the Newtonian and linear pieces. Moreover, if this overall
normalization
is fixed by the peak in the rise of the inner rotation curve (the so called
maximum disk
fit in which the Newtonian disk contribution gets to be as large as it possibly
can be),
then essentially the entire shape of the rest of the curve is fixed by just the
one parameter
(per galaxy) $\eta$. As regards this maximum disk contribution, we note that
the Newtonian
term in Eq. (23) peaks at 2.15$R_0$ with $v^2/v_0^2$ receiving a Newtonian
contribution of
0.387. This Newtonian contribution comes down to half of this value (i.e.
0.194) at 6.03$R_0$.
Since the linear contribution is essentially negligible at 2.15$R_0$
(especially after we take
the square root to get the velocity itself), if we choose the linear
contribution at 6.03$R_0$
to be equal to the Newtonian contribution at that same distance (i.e. if
numerically we set
$\eta$ equal to a critical value of 0.067), we will then have essentially
achieved flatness over
the entire 2 to 6 scale length region. Now at 6 scale lengths both the
Newtonian and linear terms
are quickly approaching their asymptotic values exhibited in Eq. (22).
Consequently at close to
12 scale lengths (precisely at 11.62$R_0$)
the linear term contribution is just 0.387, the original maximum disk value at
2.15$R_0$.
Hence between 6 and 12 scale lengths the curve again shows little deviation
from flatness.
However since the Newtonian contribution at 12 scale lengths is slightly bigger
than the linear
contribution at 2 scale lengths, the net outcome is that by 12 scale lengths
the curve is
actually beginning to show a slight rise, with flatness only being achieved out
to about 10
scale lengths. Thus in general we see that by varying just one parameter we can
naturally
achieve flatness over the entire 2 to 10 scale length region, this intriguingly
being about as
large a range of scale lengths as has up till now been observed in any rotation
curve.
In order to see just how flat a curve it is in principle possible to obtain, we
have varied
$\eta$ as a free parameter. Our most favored generic case is then obtained when
$\eta$ takes the value 0.069 (i.e. essentially the critical value),
with the resulting generic rotational velocity curve being
plotted in Fig. (2). Over the range from 3 to 10 scale lengths the ratio
$v(r)/v_0$ is found
to take the values (0.666, 0.648, 0.632, 0.626, 0.628, 0.637, 0.651, 0.667) in
unit step
increases. Thus it has a spread of only $\pm 3 \%$ about a central value of
0.647 in this region.
Additionally, we find that even at 15 scale lengths the ratio $v(r)/v_0$ has
still only
increased to 0.763, a 14$\%$ increase over its value at 10 scale lengths. In
the upper
diagram in Fig. (2) we have plotted the generic $\eta=0.069$ rotation curve out
to 10 scale
lengths to show just how flat it can be. In the lower diagram in Fig. (2) we
have shown the
continuation out to 15 scale lengths where the ultimate asymptotic rise is
becoming apparent.
We have deliberately juxtaposed the two diagrams in Fig. (2) since the flatness
out to 10
scale lengths is usually taken as indicative of asymptotic flatness as well,
with such
ultimate flatness being characteristic (and even a primary motivation)
of both isothermal gas sphere dark matter models and the MOND alternative
(Milgrom (1983)). The possibility that flatness is only
an intermediate and not an asymptotic phenomenon is one of the most unusual and
distinctive
features of the conformal gravity theory. (Of course it is always possible to
build dark
matter models with non flat asymptotic properties (see e.g. van Albada et al.
(1985)) since
the dark matter theory is currently so unconstrained. However, our point here
is that the
conformal theory is the first theory in which rotation curves are actually
required to
ultimately rise, even being predicted to do so in advance
of any data). As regards other possible values for $\eta$, if $\eta$ exceeds
the critical
value of 0.067, then the curve will be flat for fewer scale lengths with the
rise setting
in earlier, while if $\eta$ is less than the critical value, the curve will
drop perceptibly
and come back to its maximum disk value at a greater distance.

As regards the generic critical value for $\eta$, we note that for a typical
galaxy with a mass of
10$^{11}$ solar masses and a 3 kpc scale length, the required value for the
galactic
$\gamma_{galaxy}~(=N\gamma_{star}$ where $\gamma_{star}$ is the typical
$\gamma$
used in the stellar potential $V(r)$ of Eq. (3)) then turns out to be of order
10$^{-29}$/cm, which, intriguingly, is of order the inverse Hubble radius.
Moreover, this
characteristic value is in fact numerically attained in the fits of Fig. (1)
for the stellar disk contribution in all of our four
chosen galaxies  (viz. $\gamma$(154)=2.5$\times$10$^{-30}$/cm,
$\gamma$(3198)=3.5$\times$10$^{-30}$/cm,
$\gamma$(2903)=7.6$\times$10$^{-30}$/cm,
$\gamma$(5907)=5.7$\times$10$^{-30}$/cm). (While this same cosmological value
is also found for
DDO154, in some other aspects (such as possessing a rotation curve which has no
observed flat
region at all) the fitting to this dwarf irregular is found
to be anomalous (see Mannheim (1993b)) presumably because the galaxy is gas
rather than star
dominated. Hence we shall only regard the three other galaxies, all regular
spirals, as typical
for the purposes of our discussion here). Thus not only
is $\gamma_{galaxy}$ making the observed representative curves flat, and not
only is it
doing so with an effectively universal value, it is doing so with a value which
is already
known to be of astrophysical significance; thereby suggesting that
$\gamma_{galaxy}$ may be
of cosmological origin, perhaps being
related to the scale at which galaxies fluctuate out of the cosmological
background.

Additionally, we note that this apparent universality for $\gamma_{galaxy}$ has
implications for
the status of the Tully-Fisher relation in our theory. Specifically, the
average velocity $v_{ave}$
(the velocity dispersion) of the critical generic curve is equal to the maximum
disk value
since the curve is flat. Thus we can set (using $N=2\pi\Sigma_0 R_0^2$ and
letting $L$ denote the
galactic luminosity)
$$v_{ave}^4=\left( {0.387N\beta \over R_0} \right)^2
=0.300\pi\Sigma_0\beta^2 L\left( {N \over L} \right)
\eqno(39)$$
\noindent
At the critical value for $\eta$
(the fits yield $\eta$(3198)=0.044, $\eta$(2903)=0.038, $\eta$(5907)=0.057) we
also can set
$$\gamma_{galaxy}={ 0.067N\beta \over R_0^2}=0.134\pi\Sigma_0\beta \eqno(40)$$
so that Eq. (39) may be rewritten as
$$v_{ave}^4=2.239 \gamma_{galaxy}\beta L\left( {N \over L} \right) \eqno(41)$$
\noindent
If we assume that all galaxies possess the same universal value for the mass to
light
ratio (our fits yield $M/L$(3198)=4.2, $M/L$(2903)=3.5, $M/L$(5907)=6.1 in
units of
$M_{\odot}/L_{B\odot}$), we then see that given a universal $\gamma_{galaxy}$,
Eq. (41) then yields
noneother than the Tully-Fisher velocity-luminosity relation. (Observationally
the Tully-Fisher
relation is not thought to hold for the stellar component of the dwarf
irregular DDO154, as may be
anticipated since DDO154 is phenomenologically found to have an anomalously
small $M/L$
ratio ($M/L$(154) takes the value 1.4 in our fits and is essentially zero in
the dark matter and
MOND fits of Begeman et al. (1991)) - since the above
discussion does not include any non-stellar component it is anyway not
applicable to gas dominated
galaxies). Additionally, according to Eq. (40) the universality of
$\gamma_{galaxy}$ also entails
the universality of $\Sigma_0$, the central surface brightness, a
phenomenological feature first
identified for spirals by Freeman (1970). (In turn the universality of
$\Sigma_0$ entails a
mass - radius squared relation for galaxies). The (near) universality of
$\gamma_{galaxy}$ and
$\eta$
thus correlates in one fell swoop the observed flatness of rotation curves, the
universality of
$\Sigma_0$, and the Tully-Fisher relation, and does so in a theory in which
rotation curves must
eventually rise. (In passing, we note that in his review article Sanders (1990)
argues against the
possibility of being able to do precisely this in a theory which possesses one
new
non-Newtonian scale (such as $\gamma_{galaxy}$). However, his arguments were
made in the explicit
context of asymptotically flat rotation curves, and are thus bypassed here
since our curves
only enjoy flatness as an intermediate phenomenon). Thus the establishing of a
cosmological origin
for $\gamma_{galaxy}$ and $\eta$ (which would establish a cosmological origin
for $N\beta/R_0^2$
as well thereby making the galactic Newtonian and linear contributions
comparable) would
then lead naturally to flatness and the Tully-Fisher relation. The above given
discussion provides
a generalization to axially symmetric systems of an earlier discussion (Kazanas
(1991), Mannheim
and Kazanas (1991b)) based on the simplification of using Eq. (2) itself as the
galactic metric. As
we now see, the ideas developed in those two earlier papers carry over to the
present more detailed
treatment. (In passing we should point out the mass - radius squared relation
which was also
identified in those two previous papers was actually found to have
phenomenological validity on many
other astrophysical scales as well, something which still awaits an
explanation).

While we have categorized our fits as having two parameters per galaxy, the
actual situation is
slightly more constrained. Specifically, we note that the Newtonian and linear
contributions are
both proportional to $N$ according to Eq. (22). Thus if there existed universal
average stellar
parameters $\beta$ and $\gamma$ to serve as input for Eqs. (4) and (11), $\eta$
would then be fixed
by the scale length $R_0$ of each galaxy, resulting in one parameter ($N$) per
galaxy fits.
Ordinarily, one thinks of $\beta$ as being the Schwarzschild radius of the Sun,
and then in the fits
the numerical value of the mass to light ratio of the galaxy is allowed to vary
freely in the
fitting, with $M/L$ ratios then being found which are actually remarkably close
to each other
(without such closeness there would be severe violations of the Tully-Fisher
relation because of the
$N/L=M/LM_{\odot}$ factor in Eqs. (39) and (41)). However, in reality each
galaxy comes with its own
particular mix of stars, both in overall population and, even more
significantly, in the spatial
distribution of the mix. Now, of course ideally we should integrate Eq. (4)
over the true stellar
distribution allowing $\beta$ to vary with position according to where the
light and heavy stars
(stars whose luminosities do not simply scale linearly with their masses) are
physically
located within the stellar disk. Instead we use an average $\beta$ (which
incidentally enables us to
derive exact formulas such as Eq. (9)). However, two galaxies with the
identical morphological mix
of stars but with different spatial distributions of those stars should each be
approximated by a
different average $\beta$, since the Newtonian potential weights different
distances unequally.
Since we do not give two galaxies of this type different average $\beta$
parameters to begin with,
we can then compensate later by giving them different mass to light ratios
(even though for this
particular example we gave them the same morphological mix). Hence we extract
out a quantity
$N\beta_{\odot}$ from the data which simulates $N_{ave}\beta_{ave}$ where
$N_{ave}$ is the true
average number of stars in the galaxy. Because of the difference between these
two ways of
defining the number of stars in a galaxy, it is not clear whether the currently
quoted mass to
light ratios as found in the fits (in essentially all theories of rotation
curve systematics)
are merely reflecting this difference or whether they are exploiting this
uncertainty to
come up with possibly unwarrantable mass to light ratios. Thus a first
principles determination of
actual values or of a range of allowed values of galactic mass to light ratios
prior to fitting
would be extremely desirable.

A precisely similar situation also obtains for the $\gamma$ dependent terms.
Again we use an
average stellar $\gamma$ and compensate for its possible average variation from
galaxy to galaxy
by allowing the galactic gamma to light ratio ($\gamma_{galaxy}/L=N\gamma/L$)
to vary
phenomenologically (i.e. we use $N\gamma$ to simulate  $N_{ave}\gamma_{ave}$
where
$N$ is determined once and for all by normalizing the data to
$N\beta_{\odot}$).
The fits to our representative galaxies are found to yield
$N\gamma/L_B$(3198)=3.9,
$N\gamma/L_B$(2903)=5.1, $N\gamma/L_B$(5907)=3.2 (in units of
$10^{-40}$/cm/$L_{B\odot}$),
values which again are remarkably close to each other and which are of a par
with the mass to light
ratios $M/L$(3198)=4.2, $M/L$(2903)=3.5, $M/L$(5907)=6.1 found for the same
galaxies. We would not
expect the $M/\gamma_{galaxy}$ ratio to be the same for the entire sample,
simply because even if
the stellar $\beta$ and $\gamma$ parameters were to change by the same
proportion in going from one
morphological type of star to another (a reasonable enough expectation),
nonetheless, as the
galactic spatial distributions change, the inferred average stellar $\beta$ and
$\gamma$ parameters
would then change in essentially unrelated ways, since the Newtonian
and linear potentials weight the differing spatial regions of the galaxy quite
differently to
thus yield different average values. Nonetheless, it is intriguing to find that
the variation in
the average $\beta$ and $\gamma$ shows such mild dependence on specific galaxy
within our sample;
$N$ and $N_{ave}$ thus appear to be very close. To within this (mild)
variation, our fits are
thus effectively one parameter per galaxy fits. (In its pure form the MOND
explanation of
the systematics of galactic rotation curves is also a one parameter per galaxy
theory. However, in
its successful practical applications (Begeman et al. (1991)), it is generally
found necessary to
introduce at least one more fitting parameter per galaxy, such as by allowing a
(generally quite
mild) variation in the fundamental acceleration parameter $a_0$ over the
galactic sample.
Phenomenologically then MOND would thus appear to be on a par with our linear
potential theory).
For our linear potential theory we note that given the apparent uniformity of
the average stellar
$\gamma/\beta$ ratio, we see that we really have to normalize $N$ to the
maximum disk mass and that
we are really not free to vary the normalizations of the Newtonian and linear
pieces separately,
since they both are proportional to $N$. Specifically, if we make the
Newtonian piece too small we would have to arbitrarily increase the linear
contribution, something
we are not able to do in a consistent manner. Thus the Newtonian contribution
in our fit cannot be
too small. Similarly, it can never be allowed to be too large (this would give
too high a velocity);
and, hence, the Newtonian contribution in our theory is bounded both above and
below, and
essentially forced to the maximum disk mass; and thus our theory is reduced to
almost parameter
free fitting. Since dark matter fits can generally adjust the relative
strengths of the luminous
and dark matter pieces at will, they are not so constrained, and often
yield much smaller luminous Newtonian contributions, and thus large amounts of
dark matter. Thus a
first principles determination of galactic mass to light ratios might enable
one to discriminate
between rival theories. (Actually, the Tully-Fisher relation in the form of Eq.
(39) is also a
statement about the rotational velocity at the Newtonian inner region peak.
Since the Tully-Fisher
relation is generally found to hold phenenologically for the maximum disk
velocity (the relation
was initially found for the inner region velocity maximum long before outer
region flatness was
ever established), that fact alone would seem to constrain the mass to light
ratios to the values
implied by Eq. (39); to thus potentially constrain the dark matter models,
models which generally
seem to be curiously silent regarding the whole issue of the validity or
otherwise of the
Tully-Fisher relation).

In order to compare our work with that of other approaches it is useful
to clarify the significance of the term 'flat rotation curve'. In the
literature it is generally thought that rotation curves will be flat
asymptotically (though of course the more significant issue here is the fact
that the curves
deviate from the luminous Newtonian prediction at all, rather than in what
particular way);
and of course since our model predicts that velocities will
eventually grow as $r^{1/2}$, the initial expectation is that our model
is immediately ruled out. However, the rotation curve fits that have so far
been made are
not in fact asymptotic ones. Firstly, the $HII$ optical studies pioneered by
Rubin
and coworkers (Rubin et al. (1978, 1980, 1982, 1985)), even while they were
indeed yielding
flat rotation curves, were restricted to the somewhat closer in optical disk
region since the $HII$ regions are only to be found in the vicinity of hot
stars which ionize those
regions. And eventually, after a
concentrated effort to carefully measure the surface brightness of such
galaxies, it was
gradually realized
(see e.g. Kaljnas (1983) and Kent (1986)) that the $HII$ curves could be
described, albeit
coincidentally, by a standard luminous Newtonian prediction after all; even in
fact for galaxies
such as UGC2885 for which the data go out to as much as 80 kpc, a distance
which turns out to only
be of order 4 scale lengths ($R_0$=22 kpc for UGC2885, an atypically high value
- this galaxy is
just very big). Thus, not only are the optical
studies limited (by their very nature in fact) to the optical disk region where
there is some
detectable surface brightness, but it turns out, coincidentally, that they are
also limited to the
region where an extended Newtonian source is actually yielding flat rotation
curves to a rather
good degree. Thus this inner region flatness has nothing at all to do with any
possible asymptotic
flatness, though it will enable flatness to set in as early as 2 or 3 scale
lengths in fits
to any data which do go out to many more scale lengths.

While the $HII$ data do not show any substantive non-canonical behaviour,
nonetheless,
the pioneering work of Rubin and coworkers brought the whole issue of galactic
rotation
curves into prominence and stimulated a great deal of study in the field. Now
it turns
out that neutral hydrogen gas is distributed in galaxies out to much farther
distances than
the stars, thus making the $HI$ studies ideal probes of the outer reaches of
the rotation
curves and of the luminous Newtonian prediction. (That $HI$ studies might lead
to a conflict with
the luminous Newtonian prediction was noted very early by Freeman (1970) from
an analysis of
NGC300 and M33, by Roberts and Whitehurst (1975) from an analysis of M31, and
by Bosma (1978)
who made the first large 21 cm line survey of spiral galaxies).
Thus with the $HI$ studies (there are now about 30 well studied cases) it
became clear that there really was a problem with the interpretation of
galactic rotation
curve data, which the community immediately sought to explain by the
introduction of
galactic dark matter since the Newton-Einstein theory was
presumed to be beyond question. Fits to the $HI$ data have been obtained using
dark matter
(Kent (1987) provides a very complete analysis),
and while the fits are certainly phenomenologically acceptable, they
nonetheless possess
certain shortcomings. Far and away their most serious shortcoming is their ad
hoc nature,
with any found Newtonian shortfall then being retroactively fitted by an
appropriately
chosen dark matter distribution. In this sense dark matter is not a predictive
theory
at all but only a parametrization of the difference between observation and the
luminous
Newtonian expectation. As to possible dark matter distributions, no specific
distribution,
or explicit set of numerical parameters for a distribution, has
convincingly been derived from first principles as a consequence, say, of
galactic
dynamics or formation theory (for a recent critical review see Sanders (1990)).
(The general community would not appear to regard any specific derivation as
being all that
convincing since no distribution has been heralded as being so theoretically
secure
that any failure of the data to conform to it would necessitate the abandoning
of the Newton-Einstein theory).
Amongst the candidate dark matter distributions which have been considered in
the literature
the most popular is the distribution associated with a modified isothermal gas
sphere (a two parameter spherical matter density distribution
$\rho(r)=\rho_0/(r^2+r_0^2)$
with an overall scale $\rho_0$ and an arbitrarily introduced non-zero core
radius $r_0$
which would cause dark matter to predominate in the outer rather than the inner
region -
even though a true isothermal sphere would have zero core radius). The appeal
of the
isothermal gas sphere is that it leads to an asymptotically logarithmic
galactic
potential and hence to asymptotically flat rotation curves, i.e. it is
motivated by no less than the
very data that it is trying to explain. However, careful analysis of the
explicit dark
matter fits is instructive. Recalling that the inner region (around, say,
2$R_0$ for
definitiveness) is already flat for Newtonian reasons, the dark matter
parameters are then
adjusted so as to join on to this Newtonian
piece (hence the ad hoc core radius $r_0$) to give a continuously flat curve in
the
observed region. This matching of the luminous and dark matter pieces is for
the moment
completely fortuitous (van Albada and Sancisi (1986) have even referred to it
as a
conspiracy) and not yet explained by galactic dynamics. What is done in the
fits
is actually even a double conspiracy. Not only are the outer (10$R_0$) and the
inner (2$R_0$)
regions given the same velocity (by adjusting $\rho_0$), the intermediate
(6$R_0$)
region is adjusted through the core radius $r_0$ to ensure that the curve does
not fall and then
rise again in that region.
Hence flatness in the $r_0$ dominated region has almost nothing at all to do
with the
presumed asymptotically flat isothermal gas sphere contribution. Even worse, in
the actual fits the
dark matter contributions are found to actually still be rising at the largest
observed (10$R_0$)
distances, and thus not yet taking on their asymptotic values at all.
Hence the curves are made flat not by a flat dark matter contribution
but rather by an interplay between a rising dark matter piece and a falling
Newtonian one, with the
asymptotically flat expectation not yet actually having even been tested.
(Prospects for
pushing the data out to farther distances are not good because $HI$ surface
densities
typically fall off exponentially fast at the edge of the explored region). Thus
for the moment,
even though both available $HI$ and $HII$ type data sets are flat in their
respective domains,
each data set is flat for its own coincidental reason, and it would appear to
us that
region of true galactic
asymptotics has yet to be explored; with the observed flatness of the galactic
rotation curves
(just like the apparent flatness of total proton proton
scattering cross sections over many energy decades before an eventual rise)
perhaps only being
an intermediate rather than an asymptotic phenomenon.

Beyond these fitting questions (two dark matter parameters per galaxy is,
however, still fairly
economical), the outcome of the fitting is that galaxies are then 90$\%$ or so
non-luminous. Thus
not only is the Universe to be dominated by this so far undetected material,
the stars in
a galaxy are demoted to being only minor players, an afterthought as it were.
Since dark
matter only interacts gravitationally it is extremely difficult to detect (its
actual
detection with just the requisite flux would of course be a discovery of the
first magnitude),
and since it can be freely reparametrized as galactic data change or as new
data come on
line, it hardly qualifies as even being a falsifiable idea, the sine qua non
for a
physical theory. While some possible dark matter candidate particle
may eventually be detected, the issue is not whether the particle exists at all
- it may
exist for some wholly unrelated reason, but rather whether its associated flux
is big
enough to dominate galaxies.

Since the great appeal of Einstein gravity is its elegance and beauty, using an
approach as
ad hoc and contrived as dark matter for it almost defeats the whole
purpose, and would even appear to be at odds with Einstein's own view of the
way nature works.
Indeed, Einstein always referred to the Einstein Equations as being a bridge
between
the beautiful geometry of the Einstein tensor and the ugliness of the
energy-momentum tensor.
The dark matter idea only serves to make the energy-momentum tensor even more
ugly. The great
aesthetic appeal of the conformal theory is that it adds beauty to both sides
of the
gravitational equations of motion by both retaining covariance and by endowing
both the sides
of the bridge with the additional, highly restrictive, symmetry of conformal
invariance; and, as we
have seen, the theory can then even eliminate the need for dark matter
altogether. (We do not
assert that there is no dark matter in the Universe, only that there is no
gravitational basis for
assuming its existence - even with conformal gravity dark matter could still
exist for reasons
totally unrelated to gravity, albeit at a flux much lower than the standard
one).

Given the success (so far) of the linear potential theory in fitting the
rotation curve data
without needing to invoke dark matter, it would thus appear to us that at the
present time one
cannot categorically assert that the sole gravitational potential on all
distance scales is the
Newtonian one; and that, in the linear potential, the standard $1/r$ potential
would not only
appear to have a companion but to have one which would even dominate over it
asymptotically.
Indeed, the very need for dark matter in the standard theory may simply be due
to trying to apply
just the straightforward Newtonian potential in a domain for which there is no
prior (or even
current for that matter) justification. Even though the observational
confirmation on terrestrial
to solar system distance scales of both the Newton theory and its general
relativistic Einstein
corrections technically only establishes the validity of the Newton-Einstein
theory on those scales,
nonetheless, for most workers in the field, it seems to have established the
standard theory on all
other distance scales too; despite the fact that many other theories could
potentially have the
same leading perturbative structure on a given distance scale and yet differ
radically elsewhere.
Since we have shown that the conformal theory also appears to be able to meet
the constraints of
data, one has to conclude that at the present time the Newton-Einstein theory
is only sufficient to
describe data, but not yet necessary. Indeed, it is the very absence of some
principle which would
single out the Einstein theory from amongst all other possible covariant
theories which one could
in principle at least consider which prevents the Einstein theory from yet
being a necessary
theory of gravity. In fact, in a sense, it is the absence of some underlying
principle
which would ensure its uniqueness that is the major theoretical problem for the
Einstein theory,
rather than its phenomenological inability to fit data without invoking dark
matter; with this very
lack itself actually opening the door to other contenders (Mannheim (1993c)).

Finally, as regards our actual
fitting, we see that is not in fact necessary to demand flatness in the
asymptotic region
in order to obtain flat rotation curves in the explored intermediate region.
Thus, unlike
the dark matter fits, we do not need to know the structure of the data prior to
the
fitting, or need to adapt the model to a presupposed asymptotic flatness.
Further, not
only is our linear potential theory more motivated than the dark matter models
(Eq. (3)
arises in a fundamental, fully covariant, uniquely specified theory), it
possesses one fewer free
parameter per galaxy ($\gamma$ instead
of $\rho_0$ and $r_0$). Consequently, according to the usual criteria for
evaluating
rival theories, as long as conformal gravity continues to hold up, it is to be
preferred.

The author would like to thank D. Kazanas for stimulating discussions.
This work has been supported in part by the Department of Energy under
grant No. DE-FG02-92ER40716.00.
\vfill\eject

\noindent
 %\baselineskip=0.70truecm
{\bf References}
\medskip

\noindent Barnaby, D., and Thronson, H. A. 1992, A. J. 103, 41.
\smallskip
%\noindent Barnaby, D., and Thronson, H. A. 1992b, B.A.A.S. 24, 809.
%\smallskip
\noindent Begeman, K. G. 1987, Ph. D. Thesis, Gronigen University.
\smallskip
\noindent Begeman, K. G. 1989, A. A. 223, 47.
\smallskip
\noindent Begeman, K. G., Broeils, A. H., and Sanders, R. H. 1991,
Mon. Not. R. Astron. Soc. 249, 523.
\smallskip
\noindent Bosma, A. 1978, Ph. D. Thesis, Gronigen University.
%\smallskip
%\noindent Bosma, A. 1981, A. J. 86, 1791.
\smallskip
\noindent Carignan, C., and Freeman, K. C. 1988, Ap. J. (Letters) 332, 33.
\smallskip
\noindent Carignan, C., and Beaulieu, S. 1989, Ap. J. 347, 760.
\smallskip
\noindent Casertano, S. 1983, Mon. Not. R. Astron. Soc. 203, 735.
\smallskip
\noindent Casertano, S., and van Gorkom, J. H. 1991, A. J. 101, 1231.
%\smallskip
%\noindent Christodoulou, D. M. 1991, Ap. J. 372, 471.
\smallskip
\noindent Eddington, A. S. 1922, The Mathematical Theory of Relativity,
(8th Ed. 1960; Cambridge; Cambridge University Press).
\smallskip
\noindent Freeman, K. C. 1970, Ap. J. 160, 811.
\smallskip
%\noindent Hunter, D. A., Rubin, V. C., and Gallagher III, J. S. 1986, A. J.
%%91, 1086.
%\smallskip
\noindent Kalnajs, A. J. 1983, in Internal Kinematics and Dynamics of Disk
Galaxies,
IAU Symposium No. 100, ed. E. Athanassoula (Reidel, Dordrecht), p. 87.
\smallskip
\noindent Kazanas, D. 1991, Astrophysical aspects of Weyl gravity, in
Nonlinear Problems in Relativity and Cosmology, Proceedings of the Sixth
Florida Workshop on Nonlinear Astronomy, University of Florida, October
1990. Edited by J. R. Buchler, S. L. Detweiler, and J. R. Ipser,
Annals of the New York Academy of Sciences, Vol. 631, 212.
\smallskip
\noindent Kazanas, D., and Mannheim, P. D. 1991a, Ap. J. Suppl.
Ser. 76, 431.
\smallskip
\noindent Kazanas, D., and Mannheim, P. D. 1991b, Dark matter or new physics?,
in
Proceedings of the ``After the First Three Minutes" Workshop, University
of Maryland, October 1990. A. I. P. Conference Proceedings No. 222, edited by
S. S. Holt,
C. L. Bennett, and V. Trimble, A. I. P. (N. Y.).
\smallskip
\noindent Kent, S. M. 1986, A. J. 91, 1301.
\smallskip
\noindent Kent, S. M. 1987, A. J. 93, 816.
\smallskip
\noindent Krumm, N., and Burstein, D. 1984, A. J. 89, 1319.
\smallskip
\noindent Mannheim, P. D. 1990, Gen. Rel. Grav. 22, 289.
\smallskip
\noindent Mannheim, P. D. 1992, Ap. J. 391, 429.
\smallskip
\noindent Mannheim, P. D. 1993a, Gen. Rel. Grav. (in press).
\smallskip
\noindent Mannheim, P. D. 1993b, Linear potentials and galactic rotation
curves, Ap. J. (in press).
\smallskip
\noindent Mannheim, P. D. 1993c, Open problems in classical gravity, to appear
in a special issue
of Foundations of Physics in honor of the retirement of Professor Fritz
Rohrlich, L. P. Horowitz and
A. van der Merwe, Editors, Plenum Publishing Company, N.Y.
\smallskip
\noindent Mannheim, P. D., and Kazanas, D. 1989, Ap. J. 342, 635.
\smallskip
\noindent Mannheim, P. D., and Kazanas, D. 1991, Phys. Rev. D44, 417.
\smallskip
\noindent Mannheim, P. D., and Kazanas, D. 1992, Newtonian limit of
conformal gravity and the lack of necessity of the second order Poisson
equation, UCONN-92-4, unpublished.
\smallskip
\noindent Milgrom, M. 1983, Ap. J. 270, 365; 371, 384.
\smallskip
%\noindent Ostriker, J. P., and Peebles, P. J. E. 1973, Ap. J. 186, 467.
%\smallskip
\noindent Roberts, M. S., and Whitehurst, R. N. 1975, Ap. J. 201, 327.
\smallskip
\noindent Rubin, V. C., Ford W. K., and Thonnard, N. 1978, Ap. J. (Letters)
225, L107.
\smallskip
\noindent Rubin, V. C., Ford W. K., and Thonnard, N. 1980, Ap. J. 238, 471.
\smallskip
\noindent Rubin, V. C., Ford W. K., Thonnard, N., and Burstein, D. 1982,
Ap. J. 261, 439.
\smallskip
\noindent Rubin, V. C., Burstein, D., Ford W. K., and Thonnard, N. 1985,
Ap. J. 289, 81.
\smallskip
\noindent Sanders, R. H. 1990, A. A. Rev. 2, 1.
\smallskip
\noindent Tully, R. B., and Fisher, J. R. 1977, A. A. 54, 661.
\smallskip
\noindent Toomre, A. 1963, Ap. J. 138, 385.
\smallskip
\noindent van Albada, T. S., Bahcall, J. N., Begeman, K. G., and Sancisi,
R. 1985, Ap. J. 295, 305.
\smallskip
\noindent van Albada, T. S., and Sancisi, R.
1986, Phil. Trans. R. Soc. A320, 447.
\smallskip
\noindent  van der Kruit, P. C., and Searle L. 1981, A. A. 95, 105.
\smallskip
\noindent Wevers, B. M. H. R., van der Kruit, P. C., and Allen, R. J. 1986,
A. A. Suppl. Ser. 66, 505.
\medskip
\noindent
{\bf Figure Captions}
\medskip
\noindent
Figure (1). The calculated rotational velocity curves
associated with the metric of Eq. (2) for the four representative galaxies,
the intermediate sized NGC3198, the compact bright NGC2903,
the large bright NGC5907, and the dwarf irregular DDO154 (at two possible
adopted distances). In each graph
the bars show the data points with their quoted errors, the full
curve shows the overall theoretical velocity prediction (in km/s)
as a function of distance (in arc minutes) from the center of each galaxy,
while the two indicated dotted curves show the rotation curves that the
separate Newtonian and linear potentials of Eq. (3) would produce when
integrated
over the luminous matter distribution of each galaxy. No dark matter is
assumed.
\smallskip
\noindent
Figure (2). The flattest possible rotation curve for a thin exponential disk
of stars each with conformal gravity potential
$V(r)=-\beta/r+ \gamma r/2 $ which is obtained when the dimensionless ratio
$\eta$ takes the value 0.069. The full
curve shows the overall theoretical velocity prediction (in units of
$v/v_0$) as a function of distance (in units $R/R_0$),
while the two indicated dotted curves show the rotation curves that separate
Newtonian and linear potentials would produce. In the upper diagram the
rotation
curve is plotted out to 10 scale lengths to fully exhibit its flatness, while
in the lower diagram it is plotted out to 15 scale lengths to exhibit its
eventual asymptotic rise.
\end